\documentclass[10pt,conference]{IEEEtran}
\pdfoutput=1 
\usepackage{booktabs} % For formal tables
\usepackage{arydshln} % For dashed line
\PassOptionsToPackage{bookmarks=false}{hyperref}
\usepackage[T1]{fontenc}
\usepackage[latin1]{inputenc}
\usepackage[table]{xcolor}
\usepackage{enumitem}
\usepackage{graphicx}
\usepackage{multirow}
\usepackage{booktabs} % for fancy table
\usepackage{balance}
\usepackage{textcomp}
\usepackage{soul} % for highlighting text 
\usepackage{color}
\usepackage{hyperref}
\usepackage{xspace}
\usepackage{balance}
\usepackage{xcolor}
\usepackage{listings}
\usepackage{tipa}
\usepackage{url}
\ifCLASSOPTIONcompsoc
  \usepackage[nocompress]{cite}
\else
  \usepackage{cite}
\fi
\ifCLASSINFOpdf

\begin{document}
%%%%%%%%%%%%%%%%%%%%%%%%%%%%%%%%%%
\title{gazel: Supporting Source Code Edits in Eye-Tracking Studies}
%\title{An ATOM plugin for Tracking Source Code Edits during Eye-Tracking Experiments}
%\title{Supporting Source Code Edits \\ for Eye Tracking Studies }
%%%%%%%%%%%%%%%%%%%%%%%%%%%%%%%%%%

\author{
    \IEEEauthorblockA{Sarah Fakhoury, Devjeet Roy, Harry Pines, Tyler Cleveland
    }
    \IEEEauthorblockA{
    %Department of Computer Science\\
    Washington State University, USA\\
    %Washington, USA\\
    \{first.last\}@wsu.edu}
    \and
    \and
    \IEEEauthorblockA{Cole S. Peterson
    }
    \IEEEauthorblockA{%Dept. of Computer Science and Engineering \\
    University of Nebraska-Lincoln, USA\\
    %Nebraska, USA\\ 
    Cole.Scott.Peterson@huskers.unl.edu}
    \and
    \IEEEauthorblockA{Venera Arnaoudova
    }
    \IEEEauthorblockA{%Department of Computer Science\\
    Washington State University, USA\\
    %Washington, USA \\
    venera.arnaoudova@wsu.edu}
    \and
    \IEEEauthorblockA{Bonita Sharif
    }
    \IEEEauthorblockA{%Dept. of Computer Science and Engineering \\
    University of Nebraska-Lincoln, USA\\
    % Nebraska, USA\\
     bsharif@unl.edu}
    \and
    \IEEEauthorblockA{Jonathan I. Maletic
    }
    \IEEEauthorblockA{%Department of Computer Science\\
    Kent State University, USA \\
    %Ohio, USA \\
    jmaletic@kent.edu}
}
\maketitle

\newcommand{\venera}[1]{\textcolor{red}{{\it [Venera says: #1]}}} 
\newcommand{\sarah}[1]{\textcolor{blue}{{\it [Sarah says: #1]}}}
\newcommand{\devjeet}[1]{\textcolor{magenta}{{\it [Devjeet says: #1]}}}
\newcommand{\cole}[1]{\textcolor{orange}{{\it [cole says: #1]}}}
\newcommand{\bonita}[1]{\textcolor{green}{{\it [bonita says: #1]}}}
\newcommand{\johnathan}[1]{\textcolor{brown}{{\it [johnathan says: #1]}}}

\newcommand{\ic}[1]{\begin{small}\texttt{#1}\end{small}}
\def\ie{i.e.,~}
\def\eg{e.g.,~}
\def\etal{et al.~}
\def\vitalse{\ic{VITALSE}~}
\def\dejavu{\ic{D\'ej\`a Vu}~}
\def\tool{\ic{gazel}~}
%%%%%%%%%%%%%%%%%%%%%%%%%%%%%%%%%%
\begin{abstract}
Eye tracking tools are used in software engineering research to study various software development activities. However, a major limitation of these tools is their inability to track gaze data for activities that involve source code editing. We present a novel solution to support eye tracking experiments for tasks involving source code edits as an extension of the \ic{iTrace}\cite{guarnera2018itrace} community infrastructure. We introduce the \ic{iTrace-Atom} plugin and \ic{gazel} \textipa{[g@"zel]}---a Python data processing pipeline that maps gaze information to changing source code elements and provides researchers with a way to query this dynamic data. \ic{iTrace-Atom} is evaluated via a series of simulations and is over 99\% accurate at high eye-tracking speeds of over 1,000Hz. 
iTrace and \ic{gazel} completely revolutionize the way eye tracking studies are conducted in realistic settings with the presence of scrolling, context switching, and now editing. This opens the doors to support  many day-to-day software engineering tasks such as bug fixing, adding new features, and refactoring. \\

\end{abstract}
%%%%%%%%%%%%%%%%%%%%%%%%%%%%%%%%%%
\IEEEpeerreviewmaketitle
%%%%%%%%%%%%%%%%%%%%%%%%%%%%%%%%%%
\section{Introduction}

%Eye tracking source code within an integrated development environment (IDE) is a significant challenge compared to the traditional approach of using static images of source code that fit on one screen. 
Eye tracking tools and techniques are increasingly being used in software engineering research to study participants interactions with source code. 
%Several eye tracking tools exist for tracking gazes over a variety of stimuli. However, these tools lack the ability to map gazes to specific source code elements.
Traditional approaches trace gazes over static images of code, where the text does not move relative to the screen for the duration of an experiment. 
The static nature of images prevents researchers from analyzing data from experiments conducted in realistic settings, where the text visible on the screen can be dynamically changed via scrolling, editing, or switching between multiple files. 

%However, researchers often need to replicate real world scenarios with participant completing software engineering tasks, and eye tracking must happen within an IDE. 
 %In contrast to a static image, relative gaze information remains bound to a fixed \ic{(x,y)} plane, but the source code within that plane is changing. 

To address this problem, the \ic{iTrace}~\cite{guarnera2018itrace} infrastructure was developed to map gaze locations to specific source elements within an IDE. An eye tracker provides an \ic{(x,y)} coordinate indicating where a gaze is located relative to the active display, and the IDE translates this \ic{(x,y)} coordinate to a row and column in a source code file within the editor. Researchers then use parsers to identify the source code element that corresponds to the row and column being viewed. 

This process assumes that the source code file being parsed is static, \ie no edits are made during the experiment. %If the file is not static, the source element at a specific row and column is variable based on time. If source code edit information is not captured, there is no way to determine which source code element a participant was looking at the given row and column, and at what time during the experiment.
Edit support has been a widespread limitation for gaze mapping tools, and greatly impacts the types of software engineering tasks that can be studied using eye trackers (e.g., bug fix or feature addition). 
%For example, researchers have had to design program comprehension experiments that do not require the user to run or edit the source code in any way. This means that if a participant is given a bug localization task, they must read the source code and try to debug in their head, rather than write a test case, log outputs, or use the built in IDE features. Although participants are using an IDE, the software engineering process being studied is greatly limited in the types of tasks a developer might normally conduct.  There is a great need for tools that can accurately map eye tracking data to evolving source code.  

This paper presents a novel solution to address the editing limitation, as an extension to the  \ic{iTrace}~\cite{guarnera2018itrace} infrastructure. We propose \ic{iTrace-Atom}, a plugin that tracks gaze and edit information over source code files in the Atom editor, accompanied by \tool \textipa{[g@"zel]} (\underline{gaz}e \underline{e}dit evo\underline{l}ution) a Python data processing library to analyze the data collected by \ic{iTrace-Atom}.

%for software engineering researchers performing eye-tracking experiments. 
Researchers can use these tools to track source code elements as they move and change throughout the course of an experiment, all while maintaining accurate gaze fixation information. For example, researchers can track how long participants looked at a source code element of interest throughout a task, without losing valuable gaze data when an element is deleted, edited, or moved.

%Paired with \tool, our novel Atom plugin provides the following contributions to the research community: 
The paper provides the following contributions to the research community. 

\begin{itemize}
   
   \item  \textbf{Source Code Editing Support.} To enable researchers to study a variety of software tasks that involve source code editing during eye tracking experiments, we present a novel technique to capture source code edits and map eye gaze data to evolving source code. 
   
    \item \textbf{Tracking Source Code Evolution.} %Source code editing presents a unique challenge that is tracking the evolution of source code elements. For example, where and when source code elements are moved, renamed, or refactored. 
    We present a novel technique to track the evolution of source code elements, at identifier-level granularity, over the course of an experiment, including elements that are moved, renamed, or refactored. 
    
    \item  \textbf{Extended Language Support.} We present a novel data processing pipeline, \tool, in the form of an extensible Python library. The library includes parsers for all main-stream languages, improving upon existing support with 23+ new languages. 
        
    \item \textbf{Support for high-frequency eye trackers.} Eye tracking studies performed with existing plugins can drop up to 60\% of gaze information from high frequency eye trackers due to high latency from IDE plugin environment. \ic{iTrace-Atom} can capture 99\% of all gaze information from eye trackers with data sampling rates up to 2,000Hz important for in-depth cognitive analyses such as microsaccades \cite{engbert2003microsaccades}. 
    
\end{itemize}

%\bonita{In the last point above we should link the ICSME paper since we do provide high speed tracking there.  Perhaps reference it and say something like extend it to the Atom plugin. }
%Links to the Atom plugin, \tool, detailed documentation, and demonstration video can be found in our replication package \cite{replicationICSE21} \sarah{added}
\label{intro}
%%%%%%%%%%%%%%%%%%%%%%%%%%%%%%%%%%
%\section{Challenges}
%\input{Problem}
%%%%%%%%%%%%%%%%%%%%%%%%%%%%%%%%%%
\section{Approach}
\label{Methodology}

\begin{figure*}[!th]
\centering
    \includegraphics[width= .9\linewidth]{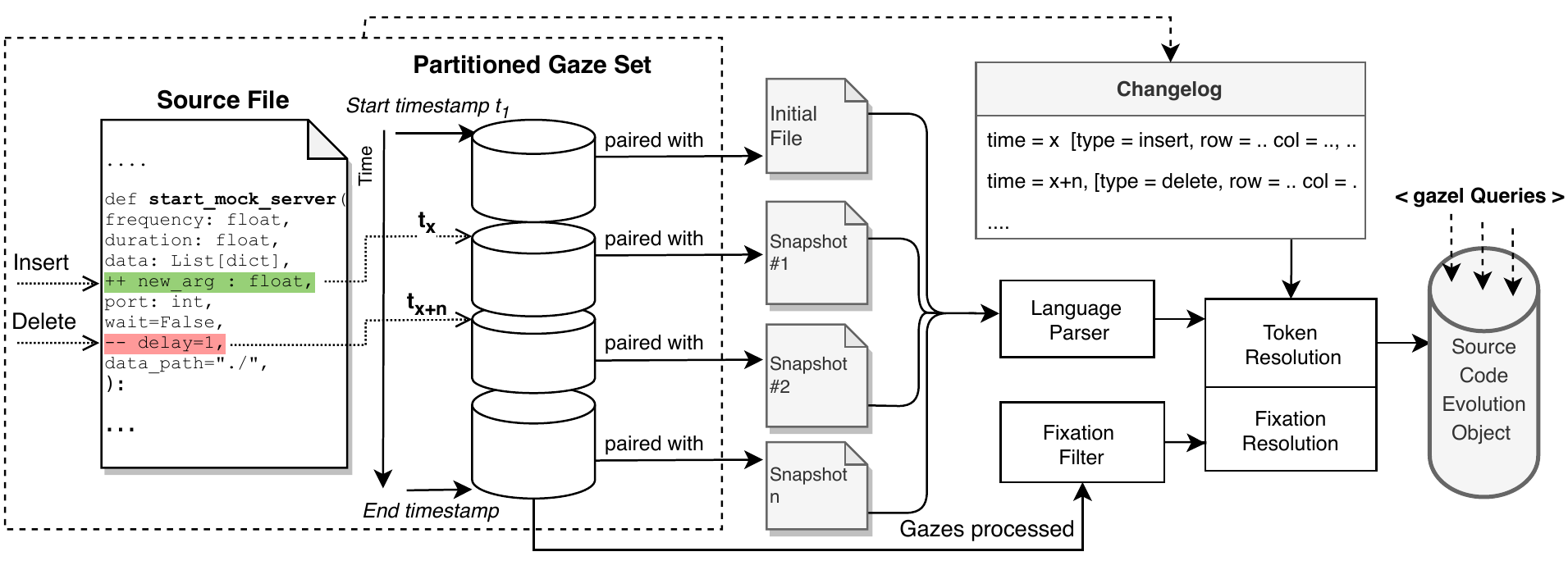}
  \caption{gazel: overview of the approach and data processing pipeline.}
  \label{fig:snapshot}
\end{figure*}

%We describe the methodology for the \ic{iTrace-Atom} plugin and data processing library, \tool, which together allow the tracking and analysis of source code edits during eye-tracking experiments.  

The steps described in Sections~\ref{sec:RecordingGazeData} and \ref{sec:CapturingEditInformation} are implemented as part of the \ic{iTrace-Atom} plugin which is published to atom.io and can installed via apm\footnote{https://atom.io/packages/itrace-atom}. The steps described in Sections~\ref{sec:CreatingSourceCodeSnapshots}--\ref{sec:TrackingChangesAcrossSnapshots} are implemented as a collection of functions part of \tool which can be installed via pypi\footnote{https://pypi.org/project/gazel/}. Documentation, detailed usage instructions, and demonstration video can be found in our online replication package\cite{replicationICSE21}.  

\subsection{Overview of the Approach}

A high level overview of the approach is shown in Figure~\ref{fig:snapshot}. 
%In the following, we describe how \tool tackles the challenges discussed in Section~\ref{sec:problem}. 
First, all edit actions performed by a user during an experiment are captured by \ic{iTrace-Atom} and saved to a change log. Throughout the experiment, the initial version of any source code file opened in the IDE is saved to memory. After an experiment is completed, \tool 
%Moving down: uses TreeSitter\footnote{https://tree-sitter.github.io/tree-sitter/}, a parser generator tool that builds editable syntax trees,  to
parses all source code files and uses the edit information saved in the change log to re-create different versions of the file in the form of source code snapshots. Snapshots represent the version of a source file at different points in time during the experiment. 

All gazes captured by the eye tracker and resolved to line and column values by \ic{iTrace-Atom} are saved to a file, which \tool then partitions. Gazes that occurred during the time for which a specific snapshot is valid will be paired to that snapshot file. Gazes are then processed using a fixation filter. %TODO do we need a [REF] for this?.

All snapshot files will now have gaze data mapped to the correct syntactic tokens. 
However, in order to understand how gaze and code edit information changes across snapshots, \tool uses the change log to create an in-memory object to represent the source code evolution during an entire experiment. Tokens are linked across snapshots, changes to a token's content or location are resolved by time, and fixation information is mapped to tokens in the appropriate snapshot. 

\subsection{Recording Gaze Data}
\label{sec:RecordingGazeData}

Gaze data is captured by the eye trackers and streamed to \ic{iTrace-Atom} via \ic{iTrace Core}\cite{guarnera2018itrace}. Gaze data is in the form of \ic{(x,y)} screen coordinates, which \ic{iTrace-Atom} must resolve to line and column numbers. 

To do this, we first scale the \ic{(x,y)} coordinates by the scale factor of the primary active display. Any coordinates outside of the active editor window bounds are recorded as invalid gazes. We then calculate the bounds of the source code within the active file, by taking into account the file gutter, 
%which is where the line number information is displayed, 
the folded code, and any active toolbars which may offset the file location. We then pass the final coordinates to Atom's \ic{getLine} and \ic{getColumn} functions, to obtain the appropriate line and column location given an \ic{(x,y)} coordinate. 
This data is written asynchronously to an .xml 'Gaze' file, which contains data for each gaze including: \ic{(x,y)} coordinates, line and column information, timestamp, and the active file name.  

\subsection{Capturing Edit Information}
\label{sec:CapturingEditInformation}
Edit information is captured directly from the Atom IDE. Once \ic{iTrace-Atom} is in the tracking phase, a change log JSON file is created containing the following information: 
%\begin{enumerate}
\textit{File:} The file which was edited. 
\textit{Type:} The type of edit. Edits can be an insert or a delete. For example, a cut-and-paste action would be registered by the IDE as a delete and then an insert. 
\textit{Offset:} an integer denoting the position of a character in the text buffer at which the edit was made. %A character offset is an integer which denotes the position of a character in the text buffer.
\textit{Text:} The contents of the edit, i.e., the inserted or deleted text. This is always a single character if the text is inserted or deleted character by character.  %If multiple characters or lines are highlighted and then deleted or replaced, the text will contain multiple characters. 
\textit{Length:} The length of the text. This is used to determine how the character offset buffer has been modified. 
\textit{Timestamp:} The time of the edit in Unix epoch time. 
\textit{Row, Column:} The starting row and column numbers in which the edit was made.
%\end{enumerate}

\subsection{Creating Source Code Snapshots}
\label{sec:CreatingSourceCodeSnapshots}

\noindent \underline{Challenge:} 
%Supporting source code editing is a non-trivial challenge. 
In order to generate correct gaze information, eye tracking data provided from the eye tracker at time $t_n$ needs to be mapped to the appropriate source code elements for the snapshot of the code that was being viewed at the same time $t_n$. All versions of the source code need to be captured and parsed in order to generate correct gaze information.

\noindent \underline{Solution:} At the start of the experiment, the original version of the source code files open in the editor are saved. Once tracking starts, all edit information is captured in a change log throughout the experiment. 
The original version of the source code file is parsed using \ic{TreeSitter}\cite{Treesitter}, a parser generator tool that builds editable syntax trees. Edits from the change log are sequentially applied to update the syntax tree. Each updated version of the syntax tree is considered as a snapshot of the original file, at a certain point in time. Consecutive edits, which are made close in time (3 seconds by default; customisable), are aggregated and applied together, to make a single snapshot with multiple changes to the syntax tree. 

\underline{Evaluation:} To ensure correctness of the reproduction of edits in \tool several different scenarios have been thoroughly tested, including simple edits, multi-line deletes, copy-pasting, replacing of text, multi-location edits (using the multiple cursors and find/replace), and changes introduced by Atom's autocomplete functionality. To verify the correctness of snapshot reproduction \tool compares the final generated snapshot to the final file version saved by Atom plugin.

%In order to ensure the validity of snapshots, the final state of all source code files that were open in the editor are saved at the end of an experiment. Each edited file snapshot is compared using the file's final state, to ensure edits were correctly applied. 

\subsection{Partitioning the Gaze Set}
Each snapshot represents the state of the source code file at a certain point in time, between two aggregate edit actions. All gaze information that is captured between two edit actions can now be accurately mapped to source code elements in the corresponding file snapshot. 
Because the IDE resolves line and column information in real time, all line and column values recorded in the gaze file are correct. The appropriate version of the code that a user was viewing is associated with the gazes captured for that time period.

\subsection{Gaze Fixation Filters}
\tool provides the option of processing raw gaze data saved by the Atom plugin using various fixation filters to generate fixations from the combined raw gaze files tagged with AST information. Currently, \tool provides the same fixation filters from the \ic{iTrace Toolkit}\cite{toolkit}. %: a basic time-based filter, IV-T and ID-T. \sarah{cant find citations for those} 
The algorithm parameters are customizable, and options to provide alternative fixation algorithms are supported. 

\begin{figure}[!h]
\centering
    \includegraphics[width=.8\linewidth]{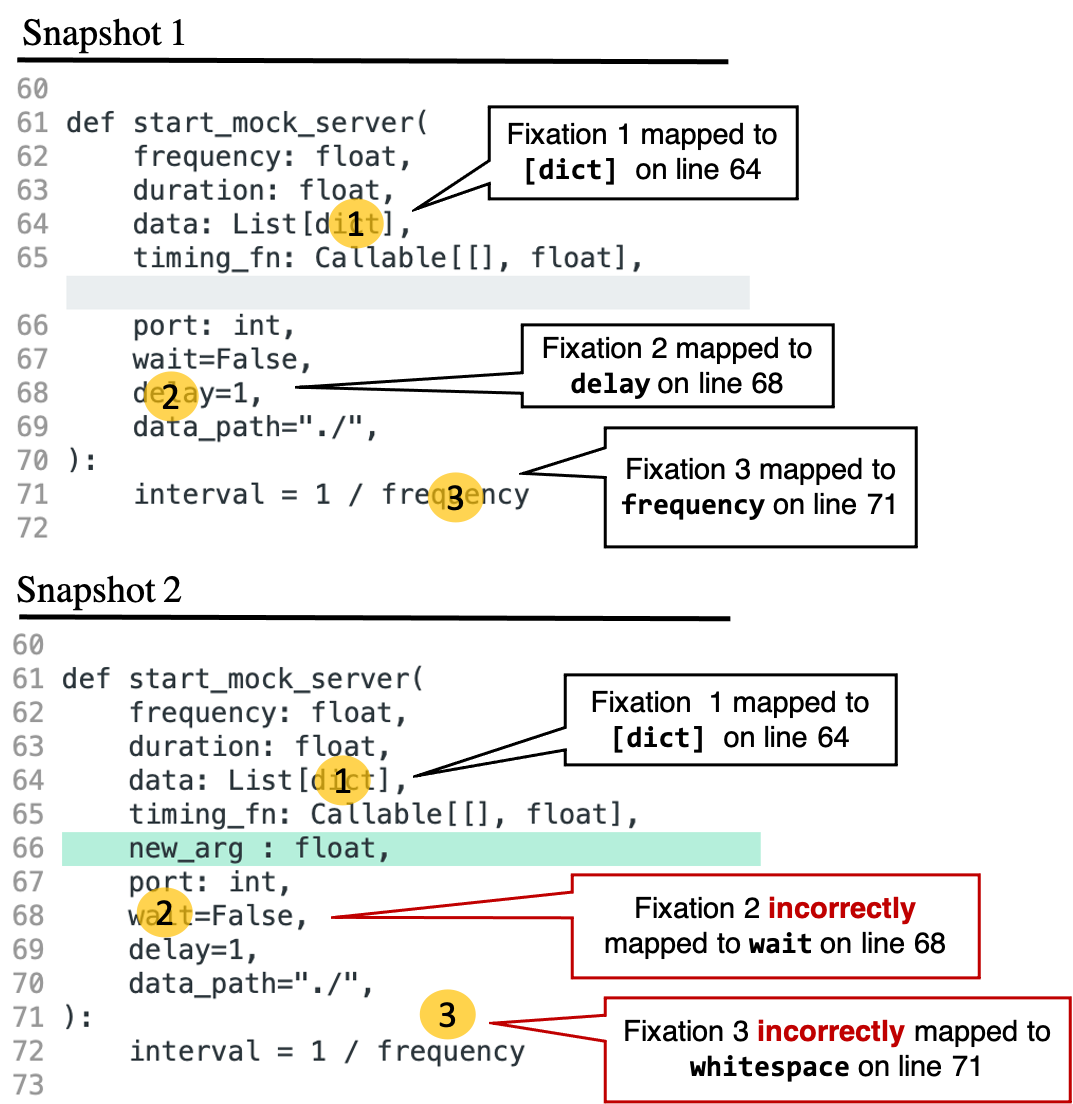}
  \caption{Challenges of tracking gazes on edited source code.}
  \label{fig:edit-challenge}
\end{figure}

\subsection{Tracking Changes Across Snapshots}
\label{sec:TrackingChangesAcrossSnapshots}
\underline{Challenge:} Tracking how gazes change as syntactic elements are edited, inserted, or deleted from a file is a major challenge.  Figure~\ref{fig:edit-challenge} contains two snapshots of a source code file, \ic{snapshot 1} is the original file and \ic{snapshot 2} has an new line inserted on line 66. Three gaze fixations recorded on \ic{snapshot 1} are displayed as yellow circles on the figure. Fixation 1 is recorded on line 64, fixation 2 on line 68, and fixation 3 on line 71. 
Current solutions assume the text at a specific line and column is static, and resolve syntactic tokens by parsing the source file. However, when a new line is inserted on line 66, fixation 2 and 3 are still mapped to lines 68 and 71. However, the syntactic token under the fixation has changed. Fixation 3 was correctly mapped on to the identifier 'frequency,' but with the edit, is now mapped to white space. A viable solution needs to keep track of all edits, and resolve gazes the correct syntactic tokens.

\underline{Solution:}
\tool provides functionality to track both source code elements and gazes across edits. \tool goes over every edit in the change log, and detects which tokens and gazes existing before the edit are affected by it. This allows tracking a token from the original version of the file to the final version with a record of how it changes, along with any fixations associated with it. 

This is implemented in a two-step approach: First, for the original version of the source code, we construct a parse tree in which we assign a unique id to every token; we update fixation data with the source code token information as well as the ids.
Second, for every edit, we apply it and we parse the source code. We then use \ic{Tree-Sitter} to obtain all the tokens in the parse tree that semantically changed from the previous version of the parse tree.
If a token has not changed semantically, we assign it the same id that it had in the previous version of the source. Otherwise, the token is assigned a new unique id. We then update fixation data with the source code token information as well as the ids.

% \begin{enumerate}
%     \item 
%         \begin{enumerate}
%             \item construct a parse tree of the source
%             \item assign a unique id to every token in the parse tree
%             \item update fixation data with the source code token information as well as the ids.
%         \end{enumerate}
%     \item Now, for every edit:
%         \begin{enumerate}
%             \item parse the source code after applying the edit
%             \item use Tree-Sitter to obtain all the tokens in the parse tree that semantically changed from the previous version of the parse tree
%             \item for any token that has not changed semantically, assign it the same id that it had in the previous version of the source. 
%             \item for any token that semantically changed as a result of the edit, assign it a new unique id.
%             \item update fixation data with the source code token information as well as the ids.
%         \end{enumerate}
% \end{enumerate}

At the end of this process, each source code element in each version of the file will be tagged with an id. This id is stable through time; if two source code elements have the same id, they are guaranteed to be the same source code element regardless of time or source file version. It is important to note here that fixation data is always correct for the snapshot of the file at the time it was recorded, and hence can be reliably attributed to a source code token if applicable. Thus, we only need to track source code elements across edits, and that information can then be used to recover the gazes on a specific source code element from previous snapshots. 
\tool provides a high level API to allow the use this information to perform aggregate analysis. For example, it allows users to obtain all fixations over a single or a set of source code elements across multiple edits, get fixation data with adjusted position information for a given snapshot (see Figure~\ref{fig:edit-challenge}), and identify what tokens changed due to a particular edit, and how it affected their associated gazes.

\subsection{Supporting High Speed Eye Trackers}
\underline{Challenge:}
Existing eye tracking plugins struggle with the high latency of IDE API calls, which prevents real time gaze and textual analysis at the data sampling rate of high-speed eye trackers as gazes are lost due to the high latency.
%This is because IDEs like Visual Studio, do not expose direct APIs for solving line and column indices and when data sampling rates exceed the API call time, gazes from the eyetracker are not captured by the plugin. 
%Support for high speed eye trackers is crucial for software engineering researchers to use different metrics of gaze information, such as saccades, and improved precision and accuracy.
One way that researchers have addressed this problem is to move any real-time analysis into an offline tool (\eg \dejavu \cite{dejavu}). However, software engineering researchers often require quick analysis of eye tracking data for use in post-experiment walk-through of the data to gain more insight into the gathered data  %comprehension processes or unexpected behavioral patterns. Quick 
for which real-time resolution of \ic{(x,y)} coordinates is needed.

\underline{Solution:}
To support real-time resolution of \ic{(x,y)} coordinates, we create \ic{iTrace-Atom}---an open-source cross platform text editor. Atom exposes APIs to efficiently map gaze data to line and columns, and is extremely fast, taking less than $1ms$ to resolve \ic{(x,y)} coordinates to line and column in the text editor (compared to existing plugins for Visual Studio where a single API call takes $15ms$). 
To ensure no gazes are dropped, we limit the number of API calls needed at the eye tracking data rate by caching %editor information, 
and writing 
%edit and gaze 
data asynchronously.  

\underline{Evaluation:} To evaluate the efficiency of \ic{iTrace-Atom}, we run an experiment by generating mock eye tracking data within the editor window at rates from 60Hz--2,000Hz for variable lengths of experiment time, and number of files open in the editor. We compare the number of gazes sent by the script to the number of gazes received and resolved to line and column numbers by the plugin. For data rates 60Hz--120Hz Atom is able to capture 100\% of gazes. For 150Hz--2,000Hz Atom's data retention drops slightly but remains higher than 99.7\%. Number of files and experiment time has no measurable effect.
Compared to existing plugins this is a major improvement. \ic{iTrace-VisualStudio} plugin captures 64\% of gazes at 60Hz and sharply drops to 6\% and 3\% for rates of 1,000Hz and 2,000Hz.  This decreases even more when multiple files are open in the editor~\cite{dejavu}.
\ic{iTrace-Eclipse} plugin performs similarly to \ic{iTrace-Atom} up to 300Hz, but then can only capture 30--15\% of gazes at 1,000--2,000Hz.
Experiment data and scripts can be found in our replication package \cite{replicationICSE21}.

%%%%%%%%%%%%%%%%%%%%%%%%%%%%%%%%%%
%\section{Implementation and Usage}
%\input{Implementation}
%\label{implementation}
%%%%%%%%%%%%%%%%%%%%%%%%%%%%%%%%%%
\section{Usage scenarios}
\label{scenarios}

%The number eye tracking studies in software engineering has been sharply increasing over the past decade, and will continue to gain popularity. 
\ic{iTrace-Atom} and \tool allow researchers to conduct a much wider variety of eye tracking studies including:
\begin{enumerate}
    \item Conducting studies involving a large variety of software maintenance tasks that involve source code editing (\eg feature implementation and testing).
    \item Analyzing the evolution of source code elements throughout the course of an experiment.
    \item Maintaining access to gaze areas of interest (AOIs) that are connected to a single source code element, even if it is modified or shifts location within a file.
%\item Collecting gaze information of novices while using a new language feature or programming construct.  
    
\end{enumerate}

\section{Related Work}

The software engineering research community has made significant effort to develop the tools and infrastructure needed to make experiments with source code and eye tracking devices feasible, with plugins designed for popular IDEs such as Visual Studio and Eclipse\cite{guarnera2018itrace, dejavu}. Researchers have also developed essential tools to help with eye tracking data processing\cite{eyecode, pygaze, pandasEye} and visualization of eye-tracking data (such as fixations and gaze paths) over source code elements \cite{roy2020vitalse, clark2017itracevis, peitek2019codersmuse}.  However, to the best of our knowledge, \tool is the first to support source code editing actions and to resolve eye-tracking data over evolving source code. \tool empowers researchers to conduct software engineering experiments with a larger spectrum of tasks where participants can edit source code, and provides a simple way for researchers to query complex source code evolution information for data analysis.

\label{relatedwork}
%%%%%%%%%%%%%%%%%%%%%%%%%%%%%%%%%%
\section{Limitations and Future Directions}
\ic{iTrace-Atom} does not track gazes that occur at the same time as edits, as they can not be accurately paired to one source code snapshot. Certain IDE features, such as split code windows, are not supported, and a detailed list of limitations can be found in \ic{iTrace-Atom}'s documentation \cite{replicationICSE21}.
In the future, we plan to add support to track refactoring edit actions across files. 
Moreover, to ensure the usefulness and usability of \tool, we plan to conduct user experience experiments with participants, and reach out to researchers to assess their needs and the usefulness of the data processing pipeline in \tool. Finally, we plan to extend edit support to the existing \ic{iTrace} plugins for Visual Studio and Eclipse in the near future, integrating it into the \ic{iTrace} infrastructure \cite{guarnera2018itrace}.

\label{Conclusion}
%%%%%%%%%%%%%%%%%%%%%%%%%%%%%%%%%%

\section{Acknowledgments}
This work is supported by the NSF (award CCF-1942228).

\bibliographystyle{plain}
\bibliography{main}

\begin{thebibliography}{10}

\bibitem{eyecode}
{Eye Code}.
\newblock \url{http://github.com/synesthesiam/eyecode}.
\newblock Accessed: 2020-11.

\bibitem{toolkit}
{iTrace-Toolkit}.
\newblock \url{http://www.i-trace.org/toolkit_doc_home_0-1-0/ }.
\newblock Accessed: 2020-11.

\bibitem{pandasEye}
{Pandas Eye}.
\newblock \url{http://github.com/hanav/PandasEye}.
\newblock Accessed: 2020-11.

\bibitem{pygaze}
{PyGaze}.
\newblock \url{http://www.pygaze.org/}.
\newblock Accessed: 2020-11.

\bibitem{replicationICSE21}
Replication-package.
\newblock \url{https://github.com/Smfakhoury/gazel}.

\bibitem{Treesitter}
{Tree-Sitter}.
\newblock \url{https://tree-sitter.github.io/tree-sitter/ }.
\newblock Accessed: 2020-11.

\bibitem{clark2017itracevis}
B.~Clark and B.~Sharif.
\newblock {iTraceVis}: Visualizing eye movement data within eclipse.
\newblock In {\em VISSOFT}, pages 22--32, 2017.

\bibitem{engbert2003microsaccades}
R.~Engbert and R.~Kliegl.
\newblock Microsaccades uncover the orientation of covert attention.
\newblock {\em Vision research}, 43(9):1035--1045, 2003.

\bibitem{guarnera2018itrace}
D.~Guarnera, C.~Bryant, A.~Mishra, J.~Maletic, and B.~Sharif.
\newblock {iTrace}: Eye tracking infrastructure for development environments.
\newblock In {\em ETRA}, pages 1--3, 2018.

\bibitem{peitek2019codersmuse}
N.~Peitek, S.~Apel, A.~Brechmann, C.~Parnin, and J.~Siegmund.
\newblock {CodersMUSE}: multi-modal data exploration of program-comprehension
  experiments.
\newblock In {\em ICPC}, pages 126--129. IEEE, 2019.

\bibitem{roy2020vitalse}
D.~Roy, S.~Fakhoury, and V.~Arnaoudova.
\newblock {VITALSE}: visualizing eye tracking and biometric data.
\newblock In {\em ICSE: Companion}, pages 57--60, 2020.

\bibitem{dejavu}
V.~Zyrianov, D.~Guarnera, C.~Peterson, C.~Scott, B.~Sharif, and J.~Maletic.
\newblock Automated recording and semantics-aware replaying of high-speed eye
  tracking and interaction data to support cognitive studies of software
  engineering tasks.
\newblock In {\em ICSME}, 2020.

\end{thebibliography}

\end{document}